\documentclass[prc,aps,twocolumn]{revtex4}
\usepackage{graphicx}
\usepackage{dcolumn}
\usepackage{bm}

\def\tend{\mathop{\to}}

\def\lim{\mathop{\rm {lim}}}

\begin{document}

\title{Nucleon dynamics in effective field theory of nuclear forces}
\author{Renat Kh.Gainutdinov and Aigul A.Mutygullina}
\address{
Department of Physics, Kazan State University, 18 Kremlevskaya St,
Kazan 420008, Russia }\email{Renat.Gainutdinov@ksu.ru}

\date{\today}
\begin{abstract}
Low energy nucleon dynamics in the effective field theory (EFT) of
nuclear forces is investigated by using the formalism of the
generalized quantum dynamics (GQD).  This formalism is based on a
generalized dynamical equation derived as the most general
equation of motion consistent with the current concepts of quantum
theory and allows one to extend quantum dynamics to the case where
the fundamental interaction in a quantum system is nonlocal in
time. The Schr{\"o}dinger equation follows from this equation in
the case where the interaction in a system is instantaneous. We
show that the effective theory of nuclear forces gives rise to low
energy nucleon dynamics which is governed by the generalized
dynamical equation with a nonlocal-in-time interaction operator.
This operator is shown to provide a natural parametrization of the
interaction of nucleons and can be derived order by order by using
methods of subtractive EFT's. We show that the use of the GQD for
describing nucleon dynamics allows one to formulate the EFT of
nuclear forces as a completely consistent theory with a
well-defined equation of motion which does not require
renormalization.
\end{abstract}
 \maketitle
\narrowtext

\section{Introduction}
\label{sec:level1}

Effective field theories (EFT's) are an important tool for
computing physical quantities in theories with disparate energy
scales [1-2]. Following the early work of Weinberg and others
[3-7], the EFT approach has become very popular in nuclear physics
[8,9]. To describe low energy processes involving nucleons and
pions, all operators consistent with the symmetries of QCD are
included in an effective Lagrangian. A fundamental difficulty is
that such a Lagrangian yields graphs which are divergent, and
gives rise to singular quantum mechanical potentials. To resolve
this problem one has to use some renormalization procedure which
regulates the integrals and subtracts the infinities in a
systematic way.  However, as is well known, there are not any
equations for renormalized amplitudes in subtractive EFT's. For
this reason in some EFT's (see, for example, Refs.[6,10,11])
finite cut-off regularization is used. Nevertheless, even when it
is assumed that the cut-off has a physical meaning renormalization
is required to render such theories consistent, and certain
cut-off-dependent terms have to be absorbed into theories
constants before determining them from empirical data.

Another problem of the EFT approach is that one cannot parametrize
the interaction of nucleons, by using some Lagrangian or
Hamiltonian. In fact, knowing the effective Lagrangian is not
sufficient to compute results for physical quantities. In
addition, one needs to specify a way to make the physical
predictions finite. The counterterms of renormalization  involve
additional unknown parameters. A renormalization scheme allow one
to use experimentally determined parameters instead of the above
unknown ones. Thus for computing physical quantities one needs not
only knowing the effective Lagrangian of the theory but also
knowing some experimentally determined parameters. The
nucleon-nucleon (NN) interaction cannot also be parametrized by
using the singular potential produced by the above effective
Lagrangian: These potentials do not make sense without
regularization and renormalization. On the other hand, if the EFT
approach, as it is widely believed, is able to provide a
fundamental description of the interactions of nucleons
  at low energies, one can hope that it will give rise to parametrization
of these interactions by an interaction operators being so
fundamental as the Coulomb potential that parametrizes the
interaction of charged particles in low energy QED. In the quantum
mechanics of particles interacting via the Coulomb potential,
which is an example of an effective theory, one deals with
well-defined interaction Hamiltonian and the Schr{\"o}dinger
equation governing the dynamics of the theory. This theory is
internally consistent and provides an excellent description of
atomic phenomena at low energies. It is natural to expect that the
EFT of nuclear forces is also able to provide such a fundamental
description of nuclear phenomena at low energies. In particular,
one can expect that this theory will allow one not only to
calculate scattering amplitudes but also to construct the
evolution operator describing the dynamics of nucleon systems. In
the present paper we show that recent developments in quantum
theory permit such a formulation of the effective theory of
nuclear forces.

The above problem of subtractive EFT's is the same that arises in
any quantum field theory with UV divergences: Regularization and
renormalization allow one to render the physical predictions
finite, however, it is impossible to construct a renormalized
Hamiltonian acting on the Fock space, i.e. after renormalization
the dynamics of the theory is not governed by the Schr{\"o}dinger
equation.  At the same time, in Ref.[12] it has been shown that
the Schr{\"o}dinger equation is not the most general dynamical
equation consistent with the current concepts of quantum physics,
and a more general equation of motion has been derived as a
consequence of the Feynman [13] and canonical approaches to
quantum theory. Being equivalent to the Schr{\"o}dinger equation
in the  case of instantaneous interactions , this generalized
dynamical equation permits the generalization to the case where
the dynamics of a quantum system is generated by a
nonlocal-in-time interaction. It has been shown [12] that a
generalized quantum dynamics (GQD) developed in this way provides
a new insight into the problem of UV divergences.

It should be noted that in the Hamiltonian formalism the
interaction generating the dynamics of an isolated system cannot
be nonlocal in time in principle. In fact, such a nonlocality of
the interaction should be associated with an energy dependence of
the interaction Hamiltonian, despite Hamiltonian itself is an
operator representing the total energy of the system. This
conceptual problem manifests itself in the fact that
energy-dependent Hamiltonians are not Hermitean, and, as a result,
the evolution of the system is not unitary. Such peculiarities are
typical in definitions of renormalized interaction Hamiltonians
(see, for example, Refs.[14,15]). The reason for this is quite
obvious. The Schr{\"o}dinger equation is local in time, and the
interaction Hamiltonian describes an instantaneous interaction.
Hence the introduction of a nonlocal-in-time (energy dependent)
interaction Hamiltonian into the Schr{\"o}dinger equation leads to
an internal inconsistency of the theory. A remarkable feature of
the generalized dynamical equation is that it provides the
extension of the theory to the case of such interactions.

Quantum mechanics is one of the basic ingredients of quantum field
theory (EFT). For this reason one may expect that in
nonrelativistic limit the Schr{\"o}dinger   equation can be used
as an equation of motion in the EFT of nuclear forces. However,
this is not true, since within the framework of this theory the
Schr{\"o}dinger equation does not make a since without
regularization and renormalization. Meanwhile, as has been shown
in Ref.[12], only the generalized dynamical equation must be
satisfied in general. This gives us the hope that this equation
can provide a perfectly satisfactory description of low energy
nucleon dynamics in the effective theory of nuclear forces.
 In the present paper we show
that in leading order of the EFT approach low energy nucleon
dynamics is governed by the generalized dynamical equation with a
nonlocal-in-time interaction operator. Moreover, this dynamics is
just the same as in the case of our model with a separable
interaction operator [12,16] that was used as a test model
illustrating the possibility of going beyond Hamiltonian dynamics
provided by the GQD. As we show in the present paper, in the case
of the quantum mechanics of nucleons at low energies, one has to
deal with such a non-Hamiltonian dynamics. In Sec.III this is
proved precisely in leading order of the EFT approach. We will
show that the leading order contact component of the two-nucleon
T-matrix obtained by summing bubble diagrams, coincides with the
T-matrix of the above model in the particular case where the form
factor in the separable interaction operator is of the form
$\varphi({\bf {p}})=1$. This T-matrix has the properties that are
at variance with the ordinary requirements of quantum mechanics,
and does not satisfy the Lippmann-Schwinger (LS) equation. At the
same time, this T-matrix is the solution of the generalized
interaction operator with a nonlocal-in-time interaction operator,
and this operator parametrizes the contact component of the NN
interaction in leading order.

In Sec.IV we will investigate the features of low energy nucleon
dynamics in leading order of the EFT approach. The possibility to
extend our results to higher order will be investigated in Sec.V
by using the example of the EFT of short-range forces developed by
van Kolck \cite{Kolck}. We will show that in next-to-leading
order, the dynamics of this theory is also governed by the
generalized dynamical equation with a nonlocal-in-time interaction
operator. By using this operator as an example, we will
demonstrate how one can construct order by order the interaction
operator parametrizing the NN interaction in the effective field
theory of nuclear forces.  The new possibilities that the GQD
opens for practical calculations in the EFT approach will be
discussed in Sec. VI. The advantages of this formalism become
apparent in the case where numerical calculations are needed, i.e.
in the case where the long-range components of the NN interactions
are taken into account. It will be shown that for solving the
evolution problem in this case the generalized dynamical equation
can be reduced to an integral equation that does not require
regularization and renormalization, and is so convenient for
numerical calculations as the LS equation.

\section{Generalized quantum dynamics}
\label{sec:level2}

Let us briefly review the main features of the formalism of the
GQD developed in Ref.[12]. As is well known, the basic concept of
the canonical formalism of quantum mechanics is that the theory
can be formulated in terms of vectors of a Hilbert space and
operators acting on this space. This formalism rests on the
postulates, which establish the connection between these
mathematical objects and observables and prescribe how to compute
the probability of an event. In the canonical formalism they are
used together with the dynamical postulate according to which the
time evolution of a quantum system is governed by the
Schr{\"o}dinger equation. At the same time, in the Feynman
formalism [13] quantum theory is formulated in terms of
probability amplitudes without resorting to the vectors and
operators acting on a Hilbert space. In this approach the
following assumption is used as the first basic postulate:

 The probability of an event is the absolute square of a
complex number called the probability amplitude. The joint
probability amplitude of a time-ordered sequence of events is
product of the separate probability amplitudes of each of these
events. The probability amplitude of an event which can happen in
several different ways is a sum of the probability amplitudes for
each of these ways.

According to this assumption, the probability amplitude of an
event which can happen in several different ways is a sum of
contributions from each alternative way. In particular, the
amplitude
 $<\psi_2| U(t,t_0)|\psi_1>$, being the probability of finding
 the quantum system in the state $|\psi_2>$ at time $t$, if at
 time $t_0$ it was in the state $|\psi_1>$, can be represented as a sum
of contributions from all alternative ways of realization of the
corresponding evolution process. Dividing these alternatives in
different classes, we can then analyze such a probability
amplitude in different ways. For example, subprocesses with
definite instants of the beginning and  end of the interaction in
the system can be considered as such alternatives. In this way the
amplitude $<\psi_2|U(t,t_0)|\psi_1>$  can be written in the form
\cite{R.Kh.:1999}
\begin{eqnarray}
<\psi_2| U(t,t_0)|\psi_1> = <\psi_2|\psi_1> \nonumber\\
+\int_{t_0}^t dt_2 \int_{t_0}^{t_2} dt_1 <\psi_2|\tilde
S(t_2,t_1)|\psi_1>, \label{repre}
\end{eqnarray}
where $<\psi_2|\tilde S(t_2,t_1)|\psi_1>$ is the probability
amplitude that if at time $t_1$ the system was in the state
$|\psi_1>,$ then the interaction in the system will begin at time
$t_1$ and will end at  time $t_2,$ and at this time the system
will be in the state $|\psi_2>.$ Here the interaction picture is
used.

According to the above postulate the probability amplitude
$<\psi_2|\tilde S(t_2,t_1)|\psi_1>$ can itself be represented by
the sum of amplitudes for each of the ways in which the subprocess
with completely specified instants of the beginning and end of the
interaction in a quantum system can happen. However, some
supplementary assumptions about the history of the system are
needed. In the Feynman approach it is assumed that this history
can be represented by some path in space-time. In this case the
amplitude $<\psi_2|\tilde S(t_2,t_1)|\psi_1>$ can be represented
by the sum of contributions from all paths corresponding to
processes in which the interaction begins at $t_1$ and ends at
$t_2$. If we assume also that the contribution from a single path
is an exponential whose (imarginary) phase is the classical action
for this path (the second postulate of Feynman's theory) and
substitute the expression obtained in this manner into Eq.(1), we
arrive at Feynman's sum-over-paths formula for the transitions
amplitudes. At the same time, in the formalism of the GQD the
history of a quantum system is represented by the version of the
time evolution of the system associated with completely specified
instants of the beginning and end of the interaction in the
system. Such a description of the history of a system is more
general and require no supplementary postulates like the second
postulate of the Feynman formalism. On the other hand, the
probability amplitudes $<\psi_2|\tilde S(t_2,t_1)|\psi_1>$, in
terms of which the evolution of a system is described within the
GQD, are used in the spirit  of Feynman's theory: The probability
amplitude of any event is represented as a sum of this amplitudes.
In Ref.[12] it has been shown that the use of the operator
formalism of the canonical approach allows one to derive a
relation for the amplitudes  $<\psi_2|\tilde S(t_2,t_1)|\psi_1>$
which can be regarded as an equation of motion.

By using the operator formalism, we can represent the probability
amplitudes $<\psi_2|U(t_2,t_1)|\psi_1>$ by the matrix elements of
the evolution operator, which must be unitary
\begin{equation}
U^{+}(t,t_0) U(t,t_0) = U(t,t_0) U^{+}(t,t_0) = {\bf 1},
\label{unitary}
\end{equation}
and must satisfy the composition law
\begin{equation}
U(t,t') U(t',t_0) = U(t,t_0), \quad U(t_0,t_0) ={\bf 1}.
\label{law}
\end{equation}
Meanwhile, $\tilde S(t_2,t_1)$ whose matrix elements are
$<\psi_2|\tilde S(t_2,t_1)|\psi_1>$ may be only an operator-valued
generalized function of $t_1$ and $t_2$, since only $U(t,t_0)={\bf
1}+ \int^{t}_{t_0} dt_2 \int^{t_2}_{t_0}dt_1\tilde S(t_2,t_1)$
must be an operator on the Hilbert space. Nevertheless, it is
convenient to call $\tilde S(t_2,t_1)$ an "operator" by using this
word in a generalized sense. In the case of an isolated system the
operator $\tilde S(t_2,t_1)$ can be represented in the form
\begin{equation}
\tilde S(t_2,t_1) = exp(iH_0t_2) \tilde T(t_2-t_1) exp(-iH_0 t_1).
\label{t}
\end{equation}

As has been shown in Ref.\cite{R.Kh.:1999}, for the evolution
operator $U(t,t_0)$ given by (\ref{repre}) to be unitary for any
times $t_0$ and $t$, the operator $\tilde S(t_2,t_1)$ must satisfy
the following equation:
\begin{eqnarray}
(t_2-t_1) \tilde S(t_2,t_1) = \int^{t_2}_{t_1} dt_4
\int^{t_4}_{t_1}dt_3 \nonumber \\
 \times(t_4-t_3) \tilde S(t_2,t_4) \tilde S(t_3,t_1).
\label{main}
\end{eqnarray}
A remarkable feature of this equation is that it works as a
recurrence relation, and allows one to obtain $\tilde S(t_2,t_1)$
for any $t_1$ and $t_2$, if it is known in an infinitesimal
neighborhood of the point $t_2=t_1$. Since the operators $\tilde
S(t_2,t_1)$ describe the contributions to the evolution operator
from the processes in which the interaction in the system begins
at $t_1$ and ends at $t_2$, the above means that in order to
construct the evolution operator it is sufficient to know the
contributions to this operator from the processes with
infinitesimal duration time of interaction. It is natural to
associate these processes with the fundamental interaction in the
system under study. This make it possible to use the relation
(\ref{main}) as a dynamical equation. One needs only to specify
the boundary condition determining the behavior of $\tilde
S(t_2,t_1)$ in the limit $t_2\to t_1$ and hence containing the
dynamical information about the system. Denoting the contribution
to the evolution operator from the processes associated with the
fundamental interaction by $H_{int}(t_2,t_1)$, such a boundary
condition can be written in the form
\begin{equation}
\tilde{S}(t_2,t_1) \tend\limits_{t_2\rightarrow t_1}
H_{int}(t_2,t_1) + o(\tau^{\epsilon}), \label{bound}
\end{equation}
where $\tau=t_2-t_1$. The parameter $\varepsilon$ is determined by
demanding that $H_{int}(t_2,t_1)$ must be so close to the solution
of Eq.(\ref{main}) in the limit $t_2\tend t_1$ that this equation
has a unique solution having the behavior (\ref{bound}) near the
point $t_2=t_1$. Within the GQD the operator $H_{int}(t_2,t_1)$
 plays the same role
as the interaction Hamiltonian in the ordinary formulation of
quantum theory: It generates the dynamics of a system. Being a
generalization of the interaction Hamiltonian, this operator is
called the generalized interaction operator.

The operator $H_{int}(t_2,t_1)$ describes fundamental processes,
starting from which, one can construct the evolution operator. In
this process the system in the state $|\psi_1>$ evolves freely up
to some time $t$ when, as a result of the interaction, the state
of the system is jumps abruptly  into the state $|\psi_2>$, and
then the system evolves freely again. The contribution from this
process into the evolution operator is of the form
$\delta(t_2-t_1)<\psi_2|A(t_1)|\psi_1>$, where the delta-function
is needed for this contribution to be nonzero. Thus in this case
the interaction operator should be of the form
\begin{equation}
H_{int}(t_2,t_1)=\delta(t_2-t_1)A(t_1).\label{dela}
\end{equation}
As has been shown in Ref.[12], the dynamical equation (\ref{main})
with the boundary condition given by (\ref{bound}) and
(\ref{dela}) is equivalent to the Schr{\"o}dinger equation with
the interaction Hamiltonian $H_I(t)=i/2A(t)$. Thus the dynamics
governed by Eq.(\ref{main}) is equivalent to Hamiltonian dynamics
in the case where the generalized interaction operator is of the
form
\begin{equation}
 H_{int}(t_2,t_1) = - 2i \delta(t_2-t_1)
 H_{I}(t_1).
 \label{delta}
\end{equation}
Correspondingly the interaction operator in the Schr{\"o}dinger
picture $$H^{(s)}_{int}(t_2-t_1) =
 \exp(-iH_0t_2)H_{int}(t_2,t_1) \exp(iH_0t_1)$$ has the form
\begin{equation}
H_{int}^{(s)}(\tau) = - 2i \delta(\tau) H_{I},
 \label{sdelta}
\end{equation}
where $H_I=\exp(-iH_0t)H_I(t) \exp(iH_0t)$.  In this case the
interaction generating the dynamics is instantaneous. On the other
hand, from both the physical and mathematical points of view there
are no reasons to restrict ourselves to the case where the
interaction operator is of the form (\ref{delta}). From the
mathematical point of view, the boundary condition (\ref{bound})
with the operator $ H_{int}(t_2,t_1)$ given by (\ref{delta}) is
not only possible boundary condition for Eq.(\ref{main}). From the
physical point of view, this equation is a unique consequence of
the representation (\ref{repre}) and the unitarity condition
(\ref{unitary}). The representation (\ref{repre}) in turn is a
consequence of the first Feynman postulate that, as is well known,
is formulated as a result of the analysis of the phenomenon of the
quantum interference and hence is one of the most fundamental
postulate of quantum theory. Thus Eq.(\ref{main}) is a unique
consequence of the first principles and can be considered as the
most general dynamical equation consistent with the current
concepts of quantum theory. Note, in this connection, that no new
fundamental concepts and postulates are used in the formalism of
the GQD. A novelty of this formalism consists in the fact that
some basic postulates of the Feynman and canonical approaches to
quantum theory are used in combination. This allows one to
formulate the theory in terms of the operators $ \tilde
S(t_2,t_1)$. As has been shown in Ref.[12], theory provides a more
detailed description of the dynamics of a quantum system then the
description directly in terms of the evolution operators, as in
the case of the canonical formalism, or in terms of Feynman's path
amplitudes. In the case where the interaction operator is of the
form (\ref{delta}), i.e. the interaction is instantaneous, the
Schr{\"o}dinger equation for the evolution operator
$$  i \frac {d U(t,t_0)}{d t} = H_I(t)U(t,t_0) $$
and Feynman's sum-over-paths formula follow from the
representation (\ref{repre}) and Eq.(\ref{main}). At the same
time, the dynamical equation permits the generalization to the
case where the interaction is nonlocal-in-time, i.e.  the time
durations of the interaction in the fundamental processes that
determine the dynamics of a system are not zero. In this case the
dynamics depends not only on the form of the operator $
H^{(s)}_{int}(\tau)$ but also on its dependence upon the duration
time $\tau$ of the interaction. However, as we have seen, only the
behavior $ H^{(s)}_{int}(\tau)$ in the limit $\tau\tend 0$ is
relevant: Knowing the behavior of $ H^{(s)}_{int}(\tau)$ in the
infinitesimal neighborhood of the point $\tau=0$ is sufficient to
construct the evolution operator by using Eq.(\ref{main}). Thus
within the GQD we deal with a new type of nonlocality. In fact,
the ordinary way of nonlocalization of a quantum field theory
consists in introducing a nonlocal form factors that depend on
parameters determining a scale of nonlocality. As for the operator
$ H_{int}(t_2,t_1)$, only its values in the infinitesimal
neighborhood of the point $t_2=t_1$ are relevant and hence the
scale of its nonlocality in time is infinitesimally small. Thus in
this case we deal with some quasilocal operators. This is very
important from the point of view of applications to  QFT where
nonlocalization aimed at resolving the problem of the UV
divergences leads to a loss of covariance of the theory. In
Ref.\cite{R.Kh.:2001} it has been shown that after renormalization
the dynamics of the three-dimensional theory of a neutral scalar
field interacting through a $\varphi^4$ coupling is governed by
the generalized dynamical equation (\ref{main}) with a
nonlocal-in-time interaction operator.

In order that the dynamical equation (\ref{main}) with the
boundary condition (\ref{bound}) have a unique solution, the
operator $ H_{int}(t_2,t_1)$ must be sufficiently close to its
relevant solution. This means that this operator  must satisfy the
condition
\begin{eqnarray}
(t_2-t_1) H_{int}(t_2,t_1)\tend\limits_{t_2 \tend t_1}
\int^{t_2}_{t_1} dt_4 \int^{t_4}_{t_1} dt_3 (t_4-t_3)\nonumber\\
 \times
H_{int}(t_2,t_4) H_{int}(t_3,t_1)+ o(\tau^{\epsilon+1}).
\label{bound'}
\end{eqnarray}
Note that the value of the parameter $\epsilon$ depends on the
form of the operator $ H_{int}(t_2,t_1).$ Since $\tilde
S(t_2,t_1)$ and $H_{int}(t_2,t_1)$ are only operator-valued
distributions, the mathematical meaning of the conditions
(\ref{bound}) and (\ref{bound'}) needs to be clarified. We will
assume that the condition (\ref{bound}) means that
\begin{eqnarray}
<\Psi_2|\int^{t}_{t_0} dt_2 \int^{t_2}_{t_0}dt_1\tilde S(t_2,t_1)
|\Psi_1>\tend\limits_{t\tend t_0}\nonumber\\
<\Psi_2|\int^{t}_{t_0} dt_2 \int^{t_2}_{t_0}dt_1
H_{int}(t_2,t_1)|\Psi_1>+o(\tau^{\epsilon+2}), \nonumber
\end{eqnarray}
 for any vectors $|\Psi_1>$ and $|\Psi_2>$ of the Hilbert space. The
condition (\ref{bound'}) has to be  considered in the same sense.

If $H_{int}(t_2,t_1)$ is specified, Eq.(\ref{main}) allows one to
find the operator $\tilde S(t_2,t_1).$ Formula (\ref{repre}) can
then be used to construct the evolution operator $U(t,t_0)$ and
accordingly the state vector
\begin{eqnarray}
|\psi(t)> = |\psi(t_0)> +  \int_{t_0}^t dt_2 \int_{t_0}^{t_2} dt_1
\tilde S(t_2,t_1) |\psi(t_0)> \label{psi}
\end{eqnarray}
 at any time $t.$ Thus
Eq.(\ref{main}) can be regarded as an equation of motion for
states of a quantum system. By using (\ref{repre}) and (\ref{t}),
the evolution operator can be represented in the form
\begin{eqnarray}
<n_2|U(t,t_0)|n_1>= <n_2|n_1>+\frac{i}{2\pi} \int^\infty_{-\infty}
dx\label{evolution}\nonumber\\
 \times \frac
{\exp[-i(z-E_{n_2})t] \exp[i(z-E_{n_1})t_0]}
{(z-E_{n_2})(z-E_{n_1})} \nonumber\\
 \times<n_2|T(z)|n_1>,
\end{eqnarray}
 where $z=x+iy$, $y>0$, and
\begin{equation}
<n_2|T(z)|n_1> = i \int_{0}^{\infty} d\tau \exp(iz\tau)
<n_2|\tilde T(\tau)|n_1>. \label{tt}
\end{equation}
Here $|n>$ are the eigenvectors of the free Hamiltonian $H_0$,
i.e. $H_0|n>=E_n|n>$, and $n$ stands for the entire set of
discrete and continuous variables characterizing the system in
full. From (\ref{evolution}), for the evolution operator in the
Schr{\"o}dinger picture, we get
\begin{equation}
U_s(t,0)=\frac{i}{2\pi}\int^\infty_{-\infty} dx\exp
(-izt)G(z),\label{schr}
\end{equation}
where
\begin{equation}
<n_2|G(z)|n_1>=\frac{<n_2|n_1>}{z-E_{n_1}}+
\frac{<n_2|T(z)|n_1>}{(z-E_{n_2})(z-E_{n_1})}.\label{lventa}
\end{equation}
Eq.(\ref{lventa}) is the well-known expression establishing the
connection between the evolution operator and the Green operator
$G(z)$ and can be regarded as the definition of the operator
$G(z)$.

The equation of motion (\ref{main}) is equivalent to the following
equation for the T-matrix \cite{R.Kh.:1999}:
\begin{equation}
\frac{d<n_2|T(z)|n_1> }{dz} =- \sum
\limits_{n}\frac{<n_2|T(z)|n><n|T(z)|n_1>}{(z-E_n)^2}, \label{dif}
\end{equation}
with the boundary condition
\begin{equation}
<n_2|T(z)|n_1> \tend \limits_{|z| \tend \infty} <n_2|
B(z)|n_1>+o(|z|^{-\beta}),\label{dbound}
\end{equation}
where $\beta=1+\epsilon,$ and
$$
<n_2|B(z)|n_1> = i \int_0^{\infty} d\tau \exp(iz \tau)
<n_2|\widetilde{B}(\tau)|n_1>,
$$
$\widetilde{B}(\tau)$ being an arbitrary operator that has the
following behavior in the limit $\tau\to 0$:
$$
<n_2|\widetilde{B}(\tau)|n_1> \tend \limits_{\tau \tend 0} <n_2|
H^{(s)}_{int}(\tau)|n_1>+o(\tau^{\epsilon}).$$ Note in this
connection that, while we define the operator $H_{int}(t_2,t_1)$
for any times $t_1$ and $t_2$, only its values for infinitesimal
duration times $\tau=t_2-t_1$ of interaction are relevant: Knowing
the behavior of  $H_{int}(t_2,t_1)$ in the infinitesimal
neighborhood of the point $t_2=t_1$ is sufficient to construct the
evolution operator by solving Eq.(\ref{main}). The two interaction
operator $H_{int}(t_2,t_1)$ and $H'_{int}(t_2,t_1)$ are
dynamically equivalent, provided
$$H_{int}(t_2,t_1)=H'_{int}(t_2,t_1)+o\left((t_2-t_1)^\varepsilon\right),
\quad \tau\tend 0.$$ In fact these operators lead to the same
solution of Eq.(\ref{main}), i.e. generate the same dynamics.
Correspondingly knowing the behavior of B(z) in the limit
$|z|\to\infty$ is sufficient for obtaining a unique solution of
Eq.(\ref{dif}). This behavior is uniquely determined by the
behavior of $H_{int}^{(s)}(\tau)$ in the limit $\tau\to 0$. At the
same time, the operator $H_{int}^{(s)}(\tau)$ need not be such
that the Fourier transform $\int_0 ^\infty
d\tau\exp(iz\tau)<n_2|H_{int}^{(s)}(\tau)|n_1>$ exists, since it
determine the behavior of the operator $\tilde T(\tau)$, for which
such a Fourier transform must satisfy, only in the limit $\tau\to
0$. Of course, one can use a dynamical equivalent operator for
which such a Fourier transform exists. The operators $\tilde
B(\tau)$ is an example of such operators. However, it is not
convenient to deal with such interaction operators. In fact, in
this case one has to take care of the behavior of
$H_{int}^{(s)}(\tau)$ not only in the limit $\tau\to 0$ but also
in the limit $\tau\to \infty$. Nevertheless, formally we can
construct the operator $B(z)$ for any $z$, by using the operator
$\widetilde{B}(\tau)$ that being dynamically equivalent to the
operator $H_{int}^{(s)}(\tau)$ satisfies the above requirement.

It should be noted that the T-matrix obtained by solving
Eq.(\ref{dif}) satisfies the following equation:
\begin{eqnarray}
<n_2|T(z_1)|n_1> - <n_2|T(z_2)|n_1> = \nonumber \\
=(z_2 -z_1) \sum_n
 \frac {<n_2|T(z_2)|n><n|T(z_1)|n_1>}
{(z_2-E_n)(z_1-E_n)}. \label{difer}
\end{eqnarray}
Below we will show that in some cases it is convenient to use
 this equation itself for obtaining the T-matrix. Note
also that in general the interaction operator has the following
form \cite{R.Kh./A.A.:1999}:
$$H_{int}(t_2,t_1)=-2i\delta(t_2-t_1)H_I(t_1)+H_{non}(t_2,t_1),$$
where the first term on the right-hand side of this equation
describes the instantaneous component of the interaction
generating the dynamics of a quantum system, while the term
$H_{non}(t_2,t_1)$ represents its nonlocal-in-time component.

\section{Nucleon dynamics in leading order of the EFT approach}

Let us now show that the leading order nucleon dynamics in the
effective theory of nuclear forces is governed by the generalized
dynamical equation (\ref{main}) with a nonlocal-in-time
interaction operator. For the sake of simplicity we will consider
the nucleons as identical, spinless particles (describe by a field
$\psi$) with 3-momenta $Q$ much smaller than their mass m, the
mass difference $\Delta$ to their first excited state, the pion
mass $\mu$, and the range of their interaction. Being
straightforward, generalization to spin and nonidentical particles
cannot give rise to an essential change of the dynamical situation
in a nucleon system. Thus we will consider the dynamics of
spinless particles at wave lengths which are large compared to the
range of their interaction. Such a dynamics can be described by
the EFT  of short-range forces developed by van Kolck
\cite{Kolck}. In this theory particles are described by a field
$\psi$, and the effective Lagrangian involve arbitrary complicated
operators of only $\psi$ and ifs derivatives. It is assumed these
derivatives are associated with factors of $1/m$, $1/\Delta$, or
$1/\mu$ and, therefore, that the effective Lagrangian can be
written as an expansion in $\partial/(m,\Delta,\mu)$.

The effective Lagrangian of the theory can be written as
\cite{Kolck}
\begin{widetext}
\begin{eqnarray}
{\cal L}=\Psi^+(i\partial_0+\frac{1}{2m}\nabla^2+\frac{1}{8m^3}
\nabla^4+\ldots)\Psi -\frac{1}{2}C_0\Psi^+\Psi\Psi^+\Psi
-\frac{1}{8}(C_2+C_2')
\{\Psi^+(\overrightarrow{\nabla}\nonumber\\-
\overleftarrow{\nabla})\Psi \cdot\Psi^+(\overrightarrow{\nabla}-
\overleftarrow{\nabla})\Psi
-\Psi^+\Psi\Psi^+(\overrightarrow{\nabla}-
\overleftarrow{\nabla})^2\Psi\} +\frac{1}{8}(C_2-C_2')\Psi^+\Psi
\nabla^2(\Psi^+\Psi)+\ldots,\label{lagrang}
\end{eqnarray}
\end{widetext} where the $C_{2n}'$s are parameters that depend on
the details of the dynamics of range $\sim 1/M$, where $M$
characterizes the typical scale of all higher-energy effects.
 Canonical quantization leads to familiar
Feynman rules, and the $\Psi$ propagator at four-momentum $p$ is
given by
\begin{equation}
S(p^0,{\bf p})=\frac{i}{p^0-\frac{{\bf p}^2}{2m}+\frac{{\bf
p}^4}{8m^3}+\ldots+i\varepsilon}.\label{prop}
\end{equation}
The four-$\Psi$ contact interaction is given by $-iv(p,p')$, with
\begin{equation}
v(p,p')=C_0+C_2({\bf p}^2+{\bf p'}^2)+2C_2'{\bf p}\cdot{\bf
p'}+\ldots,\label{pot}
\end{equation}
${\bf p} ({\bf p'})$ being the relative momentum of the incoming
(outgoing) particles. The problem can also be easily solved by
using the time-ordered perturbation theory, since in this case we
deal only with the particles evolving forward in time. Obviously
this way is more convenient for constructing the off-shell
T-matrix.

Let us consider the two-particle system at energy $E=\frac{{\bf
k}^2}{m}-\frac{{\bf k}^4}{4m^3}+\ldots$ in the center-of-mass
frame. The key point of the EFT approach is that the problem can
be solved by expanding in the number of derivatives at the
vertices or particle lines. In leading order one has to keep only
the first term $C_0$ in (\ref{pot}), and correspondingly only the
first two terms $p^0$ and $\frac{{\bf p}^2}{2m}$ in the propagator
(\ref{prop}). In this case the two particles evolve according to
the familiar nonrelativistic Schr{\"o}dinger propagator
\begin{equation}
G_0(z)=\int \frac{d^3k}{(2\pi)^3}\frac{|{\bf k}><{\bf
k}|}{z-\frac{{\bf k}^2}{m}+i\varepsilon}.
\end{equation}
In this order the Lagrangian can be rewritten in the form
\begin{eqnarray}
{\cal L}=\Psi^+(i\partial_0+\frac{1}{2m}\nabla^2)\Psi
-\frac{1}{2}C_0\Psi^+\Psi\Psi^+\Psi.\label{lagr}
\end{eqnarray}
Conservation of particle number reduces the two-nucleon T-matrix
to a sum of bubble diagrams. The ultraviolet divergences can all
be absorbed in the renormalized parameter $C_0^{(R)}$. Summing the
bubbles to a geometric series, one gets the T-matrix \cite{Kolck}
\begin{equation}
<{\bf p}_2|T^{(0)}(z)|{\bf
p}_1>=-\left[\frac{1}{C_0^{(R)}}+\frac{im^{
\frac{3}{2}}{\sqrt{z}}}{4\pi}\right]^{-1}.\label{nul}
\end{equation}
What is interesting the T-matrix given by (\ref{nul}) is exactly
the same as in the model [12,16]. This model provides an example
showing that the GQD allows one to extend quantum dynamics to the
case where the interaction generating the dynamics of quantum
system is nonlocal in time. Thus the model shows that the
situation where the dynamics of a quantum system is generated by
nonlocal-in-time interaction is possible in principle. From the
above it follows that this possibility is realized in low energy
nucleon dynamics. The same T-matrix can be obtained via the LS
equation  with the potential
\begin{equation}
V({\bf p}_2,{\bf p}_1)=\widetilde{C},
\end{equation}
by using some regularization and renormalization procedures. In
order to show this fact, let us consider the evolution problem for
two nonrelativistic particles in the c.m.s. Assume that the
generalized interaction operator in the Schr{\"o}dinger picture
 is of the form
\begin{eqnarray}
 <{\bf p}_2| {H}^{(s)}_{int}(\tau)|{\bf p}_1> =\varphi^*
 ({\bf p}_2)\varphi ({\bf p}_1)f(\tau),
 \end{eqnarray}
 where $f(\tau)$ is some function of the duration time $\tau$ of
 interaction, and the form factor  $\varphi ({\bf
p})$ must have the following asymptotic behavior for $|{\bf
p}|\to\infty$:
\begin{equation}
\varphi({\bf p}) \sim |{\bf p}|^{-\alpha}, \quad {(|{\bf p}| \tend
\infty)}. \label{form}
\end{equation}
If the interaction is instantaneous, then
$f(\tau)=-2i\lambda\delta(\tau)$, and
\begin{equation}
<{\bf p}_2| {H}^{(s)}_{int}(\tau)|{\bf p}_1>  = - 2i
\delta(t_2-t_1)<{\bf p}_2| V|{\bf p}_1>,
 \label{instdelta}
\end{equation}
$<{\bf p}_2| V|{\bf p}_1>$ being the separable potential
\begin{eqnarray}
<{\bf p}_2| V|{\bf p}_1>=\lambda\varphi^*
 ({\bf p}_2)\varphi ({\bf p}_1).
 \end{eqnarray}
 The solution of the LS equation with this potential is
\begin{eqnarray}
<{\bf p}_2|T(z)|{\bf p}_1>=\varphi^*
 ({\bf p}_2)\varphi ({\bf p}_1)\nonumber\\
 \left(\frac{1}{\lambda}+\int \frac{d^3k}{(2\pi)^3} \frac {|\varphi
({\bf k})|^2} {(z-E_k)} \right)^{-1}.
\end{eqnarray}
In the case $\alpha\leq\frac{1}{2}$, the potential (25) does not
make sense without renormalization, since it gives rise to UV
divergences. In fact, in this case the integral in (30) is not
convergent, and one has to use some regularization procedure.
Using the dimensional regularization we can write the following
expression for the regulated T-matrix:
\begin{eqnarray}
<{\bf p}_2|T_\varepsilon(z)|{\bf p}_1>=\varphi^*
 ({\bf p}_2)\varphi ({\bf p}_1)\nonumber\\
 \left(\frac{1}{\lambda_\varepsilon}+ \int \frac{d^{3-\varepsilon}k}
 {(2\pi)^3} \frac {|\varphi
({\bf k})|^2} {(z-E_k)} \right)^{-1}.
\end{eqnarray}
This expression can be rewritten in the form
\begin{eqnarray}
<{\bf p}_2|T_\varepsilon(z)|{\bf p}_1>=\varphi^*
 ({\bf p}_2)\varphi ({\bf p}_1)\nonumber\\
 \left(\frac{1}{C_0^{(\varepsilon)}}-z \int \frac{d^{3-\varepsilon}k}
 {(2\pi)^3} \frac {|\varphi
({\bf k})|^2} {(z-E_k)E_k} \right)^{-1},
\end{eqnarray}
where $(C_0^{(\varepsilon)})^{-1}=\lambda_\varepsilon^{-1}- \int
\frac{d^{3-\varepsilon}k}
 {(2\pi)^3} \frac {|\varphi
({\bf k})|^2} {E_k}$. Let us use the scattering length $a_0$ as an
additional experimental information that is needed for
renormalization. The parameter $C_0^{(\varepsilon)}$ in the
expression (32) for the T-matrix is related to the scattering
length by
$$
a_0=\frac{m}{4\pi}C_0^{(\varepsilon)}|\varphi(0)|^{-2}.
$$
The above means that we have to fix the value
$C_0^{(\varepsilon)}$ which we will denote by $C_0^{(R)}$. Now we
can let $\varepsilon\to 0$ in Eq.(32), and get
\begin{eqnarray}
<{\bf p}_2|T(z)|{\bf p}_1>=\varphi^*
 ({\bf p}_2)\varphi ({\bf p}_1)\nonumber\\
 \left(\frac{1}{C_0^{(R)}}-z \int \frac{d^{3}k}
 {(2\pi)^3} \frac {|\varphi
({\bf k})|^2} {(z-E_k)E_k} \right)^{-1}.
\end{eqnarray}
It is easy to see that this renormalized T-matrix is not a
solution of the LS equation with some potential because of the
slow rate of decay of the form factor $\varphi ({\bf p})$ as ${\bf
p}\to\infty$. Note in this connection that the renormalized
strength of the potential
$\lambda_\varepsilon=(C_0^{(\varepsilon)}- \int
\frac{d^{3-\varepsilon}k}
 {(2\pi)^3} \frac {|\varphi
({\bf k})|^2} {E_k})^{-1}$ tends to zero as $\varepsilon\to 0$.
This means that after renormalization the interaction Hamiltonian
is zero. This is a manifestation of the well-known fact that one
cannot construct a renormalized Hamiltonian, and there are not any
equations for renormalized amplitudes within Hamiltonian
formalism. At the same time, one can easily verify that the
T-matrix given by (33) satisfies the generalized dynamical
equation (\ref{main}). Moreover, as stated above, it coincides
with the T-matrix of the model [12,16]. In this model it is
assumed that the generalized interaction operator is of the form
(26). In this case, $<{\bf p}_2|\tilde S(t_2,t_1)|{\bf p}_1>$ can
be represented in the form
\begin{eqnarray}
<{\bf p}_2|\tilde S(t_2,t_1)|{\bf p}_1>=\varphi^*({\bf
p}_2)\varphi({\bf p}_1)\tilde s(t_2,t_1).
\end{eqnarray}
Correspondingly, the T-matrix should be of the form
\begin{equation}
  <{\bf p}_2| T(z)|{\bf p}_1> = \varphi^* ({\bf p}_2)\varphi ({\bf p}_1)
t(z),\label{separ}
\end{equation}
 From (16) it follows that the function $t(z)$
must satisfies the equation
\begin{equation}
\frac {dt(z)}{dz} = -t^2(z) \int \frac{d^3k}{(2\pi)^3} \frac
{|\varphi ({\bf k})|^2} {(z-E_k)^2}. \label{deq}
\end{equation}
This equation can be rewritten in the form
\begin{equation}
\frac {d\tilde{t}(z)}{dz} =\int \frac{d^3k}{(2\pi)^3} \frac
{|\varphi ({\bf k})|^2} {(z-E_k)^2} \label{deqtilde}
\end{equation}
with $\tilde{t}(z)=\left(t(z)\right)^{-1}$. All solutions of
Eq.(\ref{deq}) satisfying the condition
\begin{equation}
t^*(z)=t(z), \quad z\in(-\infty,0),
\end{equation}
are physically realizable [12]. Each of them corresponds to the
definite function $f(\tau)$ in the  generalized interaction
operator (26). As we have noted, only the asymptotic behavior of
the function $f(\tau)$ as $\tau\to 0$ is relevant, and formally
one can use any function that has this asymptotic behavior.
Knowing the asymptotic behavior of $H_{int}^{(s)}(\tau)$ in the
limit $\tau\to 0$ allows one to determine the asymptotic behavior
of the function $t(z)$ in the limit $z\to\infty$ (the one-to-one
correspondence between these behaviors follows from the Fourier
transform). This behavior in turn can be used as a boundary
condition for the differential equation (36). The T-matrix
obtained in this way can then be used for constructing the
evolution operator. Thus knowing the asymptotic behavior of
$H_{int}^{(s)}(\tau)$ in the limit $\tau\to 0$ uniquely determines
the dynamics of the system. On the other hand, as it follows from
Eq.(37), the asymptotic behavior of $t(z)$ is determined by the
large-momentum behavior of the form factor $\varphi({\bf p})$. In
the case $\alpha\geq\frac{1}{2}$, at it is easily seen from
Eq.(37), all solutions of Eq.(\ref{deq}) tend to nonzero constants
as $|z|\to\infty$, i.e. we have
\begin{equation}
t(z)\tend \limits_{|z| \tend \infty}\lambda.\label{lambda}
\end{equation}
This means that in this case the only possible form of the
function $f(\tau)$ is
\begin{equation}
f(\tau)=-2i\lambda\delta(\tau)+f'(\tau),
\end{equation}
where the function $f'(\tau)$ has no such a singularity at the
point $\tau=0$ as the delta-function. Thus, in the case
$\alpha>\frac{1}{2}$, the interaction operator should be of the
form
\begin{equation}
<{\bf p}_2| {H}^{(s)}_{int}(\tau)|{\bf p}_1>  =- 2i
\delta(\tau)<{\bf p}_2| V|{\bf p}_1>,
 \label{instdelta}
\end{equation}
$<{\bf p}_2| V|{\bf p}_1>$ being the separable potential
\begin{eqnarray}
<{\bf p}_2| V|{\bf p}_1> = \lambda \varphi^* ({\bf p}_2)\varphi
({\bf p}_1)
 \end{eqnarray}
 and hence the dynamics generated by this operator is equivalent
 to the dynamics governed by the Schr{\"o}dinger equation with the
 separable potential $\lambda\varphi^* ({\bf p}_2)\varphi ({\bf
 p}_1)$. Solving Eq.(\ref{deq}) with the boundary condition (\ref{lambda}), for the
 T-matrix, we easily get the expression (30), i.e.  in the case
 $\alpha>\frac{1}{2}$, the interaction in the system may be only
 instantaneous.

 Let us now consider the case $-\frac{1}{2}<\alpha\leq\frac{1}{2}$
 (the restriction $\alpha>-\frac{1}{2}$ is necessary for the
 integral in (\ref{deq}) to be convergent) where the ordinary
 separable-potential model does not make a sense without
 renormalization. In this case solutions of Eq.(\ref{deq}) tends to zero
 as $|z|\to\infty$:
$$
t(z)  \tend \limits_{|z| \tend \infty}  b_1
(-z)^{\alpha-\frac{1}{2}}+ b_2 (-z)^{2 \alpha-1} + o(|z|^{2
\alpha-1}), $$
 where $b_1 =- 4\pi cos(\alpha \pi)
m^{\alpha-\frac{3}{2}}$, and $b_2$ is some arbitrary constant.
Thus the T-matrix has the following asymptotic behavior for
$|z|\to\infty$:
\begin{eqnarray}
 <{\bf p}_2|T(z)|{\bf p}_1> \tend \limits_{|z| \tend \infty}  \varphi^*
 ({\bf p}_2)\varphi ({\bf p}_1)\nonumber\\
 \times\left(b_1
(-z)^{\alpha-\frac{1}{2}}+ b_2 (-z)^{2 \alpha-1} + o(|z|^{2
\alpha-1})\right),
\end{eqnarray}
By using asymptotic methods (see, for example, Ref.\cite{asym}) it
is easy to show that this large $z$ behavior of the function
$t(z)$ corresponds to the following behavior of $\tilde{T}(\tau)$
related to the T-matrix by Eq.(13) in the limit $\tau\to 0$:
\begin{eqnarray}
 <{\bf p}_2| \widetilde{T}(\tau)|{\bf p}_1> \tend \limits_{\tau\tend 0}\varphi^*
 ({\bf p}_2)\varphi ({\bf p}_1)\nonumber\\
\times\left(a_1 \tau^{-\alpha-\frac{1}{2}}
 + a_2 \tau^{-2 \alpha}+o(\tau^{-2 \alpha})\right),\label{tio}
 \end{eqnarray}
where $a_1= -ib_1 \Gamma ^{-1}(\frac{1}{2}-\alpha)
exp[i(-\frac{\alpha}{2}+ \frac{1}{4}) \pi],$ and $a_2= b_2 \Gamma
^{-1}(1-2\alpha) exp(-i \alpha \pi),$  $\Gamma(z)$ being the
gamma-function. On the other hand, according to Eqs.(4) and (6)
the asymptotic behavior of $<{\bf p}_2| \widetilde{T}(\tau)|{\bf
p}_1>$ is determined by the interaction operator
$H_{int}^{(s)}(\tau)$
\begin{equation}
<{\bf p}_2| \widetilde{T}(\tau)|{\bf p}_1> \tend\limits_{\tau\tend
0} <{\bf p}_2| H_{int}^{(s)}(\tau)|{\bf p}_1> +
o(\tau^{\epsilon}),
\end{equation}
where the parameter $\varepsilon$ is determined by demanding that
$\exp(iH_0t_2) H_{int}^{(s)}(\tau)\exp(-iH_0t_1)$ must be so close
to the relevant solution of Eq.(\ref{main}) in the limit $\tau\to
0$ that this solution is unique having the asymptotic behavior
(45).  It is easy to see that in the separable case all solutions
of Eq.(\ref{deq}) for given $\alpha$ have the same leading term in
the expansion (43), since the parameter $b_1$ is uniquely
determined by the value of the parameter $\alpha$. Only the second
term containing free parameter $b_2$ distinguishes the different
solutions of this equation. On the other hand, the first two term
in the expansion (44 ) are uniquely determined by the first two
term of the expansion (43). This means that the parameter
$\varepsilon$ in (45) is equal to $2\alpha$, and hence the
generalized interaction operator $H_{int}^{(s)}(\tau)$ should be
of the form
\begin{eqnarray}
 <{\bf p}'|H_{int}^{(s)}(\tau)|{\bf p}> =\varphi^*
 ({\bf p}')\varphi ({\bf p})
\left(a_1 \tau^{-\alpha-\frac{1}{2}}
 + a_2 \tau^{-2 \alpha}\right).\label{hio}
 \end{eqnarray}
Of course, the interaction operator (\ref{hio}) may be
supplemented by any term being of order $o(\tau^{-2 \alpha})$.
However, this will not change the solution of Eq.(\ref{main}),
i.e. in the is case we will get the generalized interaction
operator being dynamically equivalent to the interaction operator
({\ref{hio}). As we have noted, the operator $H_{int}^{(s)}(\tau)$
describes the contributions to the evolution operator from the
processes with infinitesimal duration time $\tau$ of interaction,
and only its behavior in the limit $\tau\to 0$ is relevant for us.

The solution of Eq.(37) with the initial condition
$\tilde{t}(z=0)=\tilde{t}_0$ is
\begin{equation}
\tilde{t}(z) =\tilde{t}_0-z\int \frac{d^3k}{(2\pi)^3} \frac
{|\varphi ({\bf k})|^2} {(z-E_k)E_k}. \label{soltilde}
\end{equation}
From this we easily get the expression (33) for the T-matrix. This
solution of Eq.(16) has the behavior (43), provided $
\tilde{t}_0=-b_2b_1^{-2}$. Taking this fact into account, one can
rewrite this solution in terms of the parameters $b_1$ and $ b_2$
\begin{equation}
<{\bf p}_2| T(z)|{\bf p}_1> = N(z) \varphi^* ({\bf p}_2)\varphi
({\bf p}_1),\label{nz}
\end{equation}
where
\begin{equation}
N(z)=-\left(\frac{b_2}{b_1^2}-\frac{(-z)^{\frac{1}{2}-\alpha}}{b_1}
-M(z)\right)^{-1},
\end{equation}
with
\begin{equation}
M(z) = \int\frac{d^3k}{(2\pi)^3} \frac {|\varphi({\bf k})|^2-
 |{\bf {k}}|^{-2\alpha}}
{z-E_k}.
\end{equation}

 By using Eqs.(12) and (47), we
can construct the evolution operator
\begin{eqnarray} <{\bf
p}_2|U(t,t_0)|{\bf p}_1> =<{\bf p}_2|{\bf p}_1> + \frac {i}{2\pi}
\int_{-\infty}^{\infty} dx \nonumber\\
\frac {\exp(-izt) N(z) \varphi^* ({\bf p}_2)\varphi ({\bf p}_1)}
{(z-E_{p_2})(z-E_{p_1})},\label{ev}
\end{eqnarray}
where $z=x+iy$, $y>0$. The evolution operator (\ref{ev}) satisfies
the composition law (\ref{law}) and is unitary provided the
parameter $b_2$ is real. Thus, in the case
$-\frac{1}{2}<\alpha<\frac{1}{2}$, the dynamics is well defined,
and all the requirements of quantum theory are satisfied.

The model under consideration  is an transparent illustration of
the possibility of the extension of quantum dynamics provided by
the GQD. From the point of view of Eq.(\ref{deq}), the restriction
to Hamiltonian dynamics is equivalent to the restriction to the
solutions that tend to some nonzero constant in the limit
$|z|\to\infty$. In this case the interaction in the system is
instantaneous, and the dynamics is Hamiltonian. At the same time,
this equation has solutions that tend to zero as $|z|\tend\infty$.
This takes place in the case $-\frac{1}{2}<\alpha\leq\frac{1}{2}$.
In this case the interaction is nonlocal in time and is
parameterized by the interaction operator (\ref{hio}). In this
case the dynamics is non-Hamiltonian.  There are no reasons to
restrict ourselves to the solutions of Eq.(\ref{deq}) which tend
to  nonzero constants, because all solutions of this equation
satisfying the condition (38) are physically realizable, and such
an extension of quantum dynamics is possible in principle. The
remarkable fact is that this possibility is realized in low energy
nucleon dynamics, and, as we have noted, the leading order
contribution to the nucleon-nucleon T-matrix is described by our
model. In fact, in the case $\psi({\bf p})=1$, for the T-matrix
given by (47), we have
\begin{eqnarray}
<{\bf p}_2|T(z)|{\bf
p}_1>=-\frac{b_1^2}{b_2-ib_1\sqrt{z}}=\nonumber\\
-\left[\frac{1}{C_0^{(R)}}+\frac{im^{
\frac{3}{2}}{\sqrt{z}}}{4\pi}\right]^{-1},
\end{eqnarray}
where $b_1=-\frac{4\pi}{m\sqrt{m}}$, and
$C_0^{(R)}=\frac{b_1^2}{b_2}$. In this case the generalized
interaction operator is of the form
\begin{eqnarray}
 <{\bf p}_2| {H}^{(s)}_{int}(\tau)|{\bf p}_1> =
\frac{4\sqrt{\pi}\exp(i\frac{3\pi}{4})}{m^{\frac{3}{2}}\sqrt{\tau}}+
 \frac{16\pi^2}{m^{3}C_0^{(R)}}. \label{pioper}
\end{eqnarray}
This operator parametrizes the leading order contact component of
the NN interaction and generates the two-nucleon dynamics that is
described by the evolution operator
\begin{eqnarray}
<{\bf p}_2|U(t,t_0)|{\bf p}_1> =<{\bf p}_2|{\bf p}_1> + \frac
{i}{2\pi}
\int_{-\infty}^{\infty} dx \nonumber\\
\frac {\exp(-izt) N(z)} {(z-E_{p_2})(z-E_{p_1})}.\label{eqev}
\end{eqnarray}
 It should be noted that the
interaction operator (\ref{pioper}) contains the renormalize
parameter $C_0^{(R)}$, while in the ordinary methods based on the
use of a renormalization procedure this parameter appears  in
renormalized amplitudes at the final stage. The generalized
dynamical equation (\ref{main}) with the interaction operator
(\ref{pioper}) is well defined and allows one to describe, in a
consistent way, the dynamics generated by the leading order
component of the NN interaction.

\section{Non-Hamiltonian character of low energy nucleon dynamics}

Let us now discuss advantages of the GQD and new possibilities
that it opens for describing low energy nucleon dynamics. As we
have seen, the T-matrix (33) obtained by solving the dynamical
equation (16) with the interaction operator (46), can be also
obtained starting with the singular potential
\begin{eqnarray}
 V({\bf p}',{\bf p})=\lambda \varphi^*({\bf p}')
\varphi({\bf p}),\quad \varphi({\bf p})\sim|{\bf
p}|^{-\alpha},\quad\alpha\leq\frac{1}{2}.\nonumber
\end{eqnarray}
 However, in this way one cannot determine a
potential that could parametrize the interaction in the system.
The above singular potential  does not make sense without
renormalization and is only of formal importance for the problem
under consideration. In fact, as we have shown, the strength of
potential, which in the regulated LS equation is adjusted to give
the correct scattering length, becomes zero after removing
regularization. All the information contained in this potential is
that the T-matrix is of the form (35) with the same form factor.
This information is not sufficient to construct the T-matrix. In
addition, one needs to use some experimental data, for example,
the scattering lengths. However, one cannot construct an
interaction Hamiltonian that could contain all this information.
The T-matrix (33) obtained in this way does not satisfy the LS
equation. Moreover, as we will show below, it has the properties
that are at variance with the Hamiltonian formalism. Thus in this
case we have only a calculation rule that allows one to compute
the T-matrix: One cannot derive any renormalized equation
containing a well-defined potential that parametrizes the
interaction in the system. This problem is the price for trying to
describe the dynamics of the system after renormalization in terms
of the Hamiltonian formalism, despite this dynamics is
non-Hamiltonian. From the more general point of view provided by
the GQD we see that the T-matrix (33) satisfies the generalized
dynamical equation (16) with the nonlocal-in-time interaction
operator (46), and this operator describes the fundamental
interaction in the system. Knowing this operator is sufficient to
construct the T-matrix and hence the evolution operator, and in
this case we deal with the well-defined equation that does not
require renormalization. As we have seen, an example of such a
dynamics is the leading order nucleon dynamics where, as a
consequence of the symetries of the underlying theory, the form
factor must satisfy the condition $\varphi({\bf p})=1$.

In order to clarify the character of the leading order nucleon
dynamics  let us examine the properties of the evolution operator
of the more general theory with the interaction operator (46). In
the Schr{\"o}dinger picture, this operator $ <{\bf p}_2|V(t)|{\bf
p}_1>\equiv<{\bf p}_2|U_s(t,0)|{\bf p}_1>$ can be rewritten in the
form
\begin{eqnarray}
<{\bf p}_2|V(t)|{\bf p}_1> =<{\bf p}_2|{\bf p}_1>
\exp(-iE_{p_2}t)\nonumber\\+ \frac {i}{2\pi}
\int_{-\infty}^{\infty} dx \frac {\exp(-izt) <{\bf p}_2|T(z)|{\bf
p}_1>} {(z-E_{p_2})(z-E_{p_1})},
\end{eqnarray}
where $<{\bf p}_2|T(z)|{\bf p}_1>$ is given by (\ref{nz}). Since
this T-matrix satisfies Eqs.(\ref{dif}) and (\ref{difer}), the
evolution operator (51) is unitary, and satisfies the composition
law (\ref{law}). Correspondingly the operators $V(t)$ constitute a
one-parameter group of unitary operators, with the group property
\begin{equation}
V(t_1+t_2) = V(t_1) V(t_2) , \quad    V(0)= {\bf 1}.
\end{equation}
Assume that this group has a self-adjoint infinitesimal generator
$H$ which in the Hamiltonian formalism is identified with the
total Hamiltonian. Then for $|\psi>\in{\cal D}(H)$ we have
\begin{equation}
\frac{V(t)|\psi> - |\psi>}{t}\tend\limits_{t \tend 0}-iH|\psi>.
\end{equation}
From this and (55) it follows that
$$H=H_0+H_I,$$
with
\begin{eqnarray}
<{\bf p}_2|H_I|{\bf p}_1>= \frac {i}{2\pi} \int_{-\infty}^{\infty}
dx \frac {z<{\bf p}_2|T(z)|{\bf p}_1>} {(z-E_{p_2})(z-E_{p_1})},
\end{eqnarray}
where $z=x+iy$, and $y>0$. Since $<{\bf p}_2|T(z)|{\bf p}_1>$ is
an analytic function of $z$ and, in the case
$\alpha\leq\frac{1}{2}$, tends to zero as $|z|\tend \infty$, from
Eq.(58) it follows that $<{\bf p}_2|H_I|{\bf p}_1>=0$ for any
${\bf p}_2$ and ${\bf p}_1$, and hence $H=H_0$. This means that,
if the infinitesimal generator of the group of the operators
$V(t)$ exists, then it coincides with the free Hamiltonian, and
the evolution operator is of the form $V(t)=\exp(-iH_0t)$. Thus,
since this, obviously, is not true, the group of the operators
$V(t)$ has no infinitesimal generator, and hence the dynamics is
not governed by the Schr{\"o}dinger equation.

It should be also noted that in the case $\alpha\leq\frac{1}{2}$,
$\tilde S(t_2,t_1)$ is not an operator on the Hilbert space. In
fact, the wave function
\begin{eqnarray}
\psi({\bf p})\equiv<{\bf p}|\psi>=
<{\bf p}|\tilde S(t_2,t_1)|\psi_1>=\nonumber \\
=\varphi^*({\bf p})\tilde s(t_2,t_1)\int \frac{d^3k}{(2\pi)^3}
\varphi({\bf k})<{\bf k}|\psi_1>
\end{eqnarray}
is not square integrable for any nonzero $|\psi_1>$, because of
the slow rate of decay of the form factor $\varphi({\bf p})$ as
$|{\bf p}|\tend\infty$.  Correspondingly, in the case $\alpha\leq
\frac{1}{2}$, the T-matrix given by (48) does not represent an
operator on the Hilbert space. However, as we have stated, in
general $\tilde S(t_2,t_1)$ may be only an operator-valued
generalized function such that the evolution operator is an
operator on the Hilbert space. Correspondingly the T-matrix need
not be an operator on the Hilbert space. It is enough that the
evolution operator given by (51) is such an operator. The T-matrix
and $\tilde S (t_2,t_1)$ satisfy these requirements not only for
$\alpha>\frac{1}{2}$ but also for
$-\frac{1}{2}<\alpha\leq\frac{1}{2}$, since the evolution operator
(51) is an operator on the Hilbert space. At the same time, in the
case $\alpha\leq\frac{1}{2}$ we go beyond Hamiltonian dynamics.

The above means that low energy nucleon dynamics in the effective
theory of nuclear forces cannot be governed by the Schr{\"o}dinger
equation and hence is non-Hamiltonian. Correspondingly the
interaction of nucleons cannot be parametrized by an interaction
Hamiltonian defined on the Hilbert space. At the same time, as has
been shown   on the example of the effective theory with the
Lagrangian (23), this dynamics is governed by the generalized
dynamical equation (\ref{main}) with a nonlocal-in-time
interaction operator. In the above theory describing the leading
order nucleon dynamics this operator is of the form (53), and two
nucleon dynamics is described by the evolution operator (54). This
evolution operator is unitary and satisfies the composition law
(\ref{law}). Thus the use of Eq.(\ref{main}) as an equation of
motion in the effective theory of nuclear forces allows one not
only to calculate scattering amplitudes but also to construct the
evolution operator describing low energy nucleon dynamics.

\section{next-to-leading order}

We have shown that in leading order the dynamics of the EFT under
consideration is governed by the generalized dynamical equation
with a nonlocal-in-time interaction operator. Let us now show that
this is true also in  next-to-leading order. For this let us come
back to the theory with the Lagrangian (19) in which it is assumed
that particles evolve forward in time, and particle number is
conserved. Summing the graphs of the perturbation series yield the
following on-shell T-matrix \cite{Kolck}:
\begin{eqnarray}
T_{os}(k,{\bf p'}\cdot{\bf p})=T_{os}^{(0)}(k)-
2{C'_2}^{(R)}k\widehat{\bf p'}\cdot\widehat{\bf
p}\nonumber\\+O[\frac{4\pi}{mM}({Q}/M)^4],
\end{eqnarray}
where $T_{os}^{(0)}(k)$ is the S-wave amplitude and is given by
\begin{eqnarray}
 T_{os}^{(0)}(k)=\left[\frac{1}{C_0^{(R)}}-2\frac{C_2^{(R)}}
 {(C_0^{(R)})^2}k^2+
 \frac{imk}{4\pi}\left(1+\frac{k^2}{2m^2}\right)\right]^{-1}\\
 \times[1+O((Q/M)^4)],\nonumber
\end{eqnarray}
$k=|{\bf p}_1|=|{\bf p}_2|,$ $\widehat{{\bf p}_i}/k$, $C_0^{(R)}$,
$C_2^{(R)}$ and ${C'_2}^{(R)}$ are renormalized parameters. In
this way one can also construct the off-shell T-matrix. At the
same time, as it has been shown in  leading order, the T-matrix
can be obtained without summing the diagrams and resorting to
regularization and renormalization procedures, provided that the
UV behavior of the T-matrix elements as functions of momenta is
known. As we have seen, the requirement that the T-matrix being of
the form $<{\bf p}_2|T(z)|{\bf p}_1>=t(z)$ satisfies the
generalized dynamical equation (16) yields the formula (52)  where
only the parameter $C_0^{(R)}$ is free. In next-to-leading order
the relation between the Lagrangian (19) and the form of the
T-matrix is not so straightforward, because in this case one has
to take into account the fact that this Lagrangian is only of the
formal importance, and should be supplemented by some
counterterms. Nevertheless, let us assume that they do not effect
on the form of the T-matrix. In this case from the analysis of the
perturbation series of the theory it follows that the two-particle
T-matrix (in this paper we focus on the S-wave channel) should be
of the form
\begin{eqnarray}
<{\bf p}_2|T(z)|{\bf p}_1>=t_1(z)+C_1t_2(z)({\bf p}_1^2+ {{\bf
p}_2}^2)+ \nonumber\\+O[\frac{4\pi}{mM}(Q/M)^4],\label{two}
\end{eqnarray}
where $C_1=C_2^{(R)}/C_0^{(R)}$. Substituting (\ref{two}) into
(16), we get
\begin{widetext}
\begin{eqnarray}
\frac {dt_1(z)}{dz} = - \int
\frac{d^3k}{(2\pi)^3}\frac{1}{(z-\frac{k^2}{m})^2}\left[t_1^2(z)\left(1+\frac{2k^4}
{4m^3(z-\frac{k^2}{m})}\right)+2C_1t_1(z)t_2(z)k^2\right]+O[\frac{4\pi}{mM}(Q/M)^4].
\label{solv}
\end{eqnarray}
\end{widetext}
Since the problem with UV divergences must not arise in treating
the generalized dynamical equation, we have
\begin{equation}
C_1=-\frac{1}{4m^2}+\delta C_1
\end{equation}
with $\delta C_1=O(\frac{1}{Mm^2})$ and
\begin{equation}
t_1(z)=t_2(z)=t(z)\left(1+O[(\frac{Q}{M})^4]\right),
\end{equation}
where $t(z)$ is a solution of the equation
\begin{eqnarray}
\frac {dt(z)}{dz} = -t^2(z)\int
\frac{d^3k}{(2\pi)^3(z-\frac{k^2}{m})^2}
\left(1+\frac{2zk^2}{2m^2(z-\frac{k^2}{m})}\right).
\end{eqnarray}
Correspondingly, the function
$\tilde{t}(z)\equiv\left(t(z)\right)^{-1}$ satisfies the following
equation:
\begin{eqnarray}
\frac {d\tilde{t}(z)}{dz} = \int
\frac{d^3k}{(2\pi)^3(z-\frac{k^2}{m})^2}
\left(1+\frac{2zk^2}{2m^2(z-\frac{k^2}{m})}\right).
\end{eqnarray}
Its solution with the initial condition
$\tilde{t}(z=0)=\tilde{t}_0$ is
\begin{widetext}
\begin{eqnarray}
\tilde{t}(z) =\tilde{t}_0+\int\limits_0^zds\int
\frac{d^3k}{(2\pi)^3} \frac
{1}{(s-\frac{k^2}{m})^2}\left(1+\frac{2sk^2}{2m^2(s-\frac{k^2}{m})}\right)\nonumber\\=
\tilde{t}_0+\int \frac{d^3k}{(2\pi)^3} \left(\frac
{z+\frac{z^2}{4m}}{(z-\frac{k^2}{m})(-\frac{k^2}{m})}-
\frac{z^3}{4m(z-\frac{k^2}{m})^2(-\frac{k^2}{m})}\right)
=\tilde{t}_0-\frac{im^{\frac{3}{2}}}{4\pi}\sqrt{z}\left(1+\frac{z}{8m}\right).
\label{solti}
\end{eqnarray}
Denoting $\tilde{t}_0=-\frac{1}{C_0^{(R)}}$, for the function
$t(z)$, we get
\begin{eqnarray}
t(z)=-\left[\frac{1}{C_0^{(R)}}+\frac
 {im^{\frac{3}{2}}}{4\pi}\sqrt{z}\left(1+\frac{z}{8m}\right)\right]^{-1}
[1+O((Q/M)^4)].
\end{eqnarray}
Substituting (65) with $t(z)$ given by (69) into (63) yields
\begin{eqnarray}
<{\bf p}_2|T(z)|{\bf p}_1> = -\frac{1+\frac{C_2{(R)}}{C_0^{(R)}}
({\bf p}_1^2+ {\bf{p}}_2^2)}{\frac{1}{C_0^{(R)}}+
 im^{3/2}(4\pi)^{-1}\sqrt{z} (1+\frac{z}{8m})}+O[\frac{4\pi}{mM}(Q/M)^4].
\end{eqnarray}
One can obtain the on-shell T-matrix, by putting $|{\bf
p}_1|=|{\bf p}_2|=k$, and $z=\frac{k^2}{m}-\frac{k^4}{4m^3}$:
\begin{eqnarray}
T_{os}(k,{\bf p'}\cdot{\bf p})
=-\left(1+2\frac{C_2^{(R)}}{C_0^{(R)}}k^2\right)
\left[\frac{1}{C_0^{(R)}}+
 \frac{im^{\frac{3}{2}}}{4\pi}\sqrt{\frac{k^2}{m}-\frac{k^4}{4m^3}}
 \left(1+\frac{k^2}{8m^2}+\frac{k^4}{32m^4}\right)\right]^{-1}\nonumber\\
+O[\frac{4\pi}{mM}(Q/M)^4]
=-\left[\frac{1}{C_0^{(R)}}-2\frac{C_2^{(R)}}{(C_0^{(R)})^2}k^2+
 \frac{imk}{4\pi}\left(1+\frac{k^2}{2m^2}\right)\right]^{-1}
 + O[\frac{4\pi}{mM}(Q/M)^4].
\end{eqnarray}
\end{widetext}
 Thus the requirement that the T-matrix of the form (\ref{two}) satisfies
the generalized dynamical equation (16) yields the formula (70),
and in this way we get the same two-particle on-shell T-matrix
that in Ref.\cite{Kolck} has been obtained by summing the bubble
graphs of the EFT.

In order to obtain the form of the generalized interaction
operator that leads to the T-matrix (70), one has to examine the
large $z$ behavior of the  function $t(z)$ in the limit
$|z|\to\infty$. From (69) it follows that
\begin{eqnarray}
t(z)\tend\limits_{|z|\tend\infty}\frac{b_1}{\sqrt{-s}}+\frac{b_2}{(-s)}+o(|s|^{-1}),
\end{eqnarray}
where $s=z\left(1+\frac{z}{4m}\right)$, $b_1=
-\frac{4\pi}{m\sqrt{m}}$, and $b_2=\frac{b_1^2}{C_0^{(R)}}$. Of
course, the asymptotic behavior of $t(z)$ can be represented in
the form of the expansion in $z^{-\frac{1}{2}}$:
\begin{eqnarray}
t(z)\tend\limits_{|z|\tend\infty}\frac{b'_1}{(-z)^{\frac{3}{2}}}+
\frac{8mb'_1}{(-z)^{\frac{5}{2}}}+\frac{b'_2}{(-z)^{3}}+o(|z|^{-3}),\label{expan}
\end{eqnarray}
where $b'_1=\frac{32\pi}{\sqrt{m}}$ and
$b'_2=\frac{(b'_1)^2}{C_0^{(R)}}$. However, for the problem under
consideration the values of z much larger than $Q$ but much
smaller than $m$ should be considered as infinitely large, and the
behavior of the function in this region is described by the
expansion (\ref{expan}). The interaction operator
$H_{int}^{(s)}(\tau)$ that gives rise  to this behavior of $t(z)$
is
\begin{eqnarray}
 <{\bf p}_2| {H}^{(s)}_{int}(\tau)|{\bf p}_1>=\left(1+\frac{C_2^{(R)}}{C_0^{(R)}}
 ({\bf p}_1^2+ {\bf{p}}_2^2)\right)\nonumber\\
\left(\frac{4\sqrt{\pi}\exp(i\frac{3\pi}{4})}{m^{\frac{3}{2}}
\sqrt{\tau-\frac{i}{8m}}}+\frac{16\pi^2}{m^3C_0^{(R)}}\right).
\label{gio}
\end{eqnarray}
This operator parametrizes the contact term of the NN interaction
in the S-wave channel up to next-to-leading order. The generalized
dynamical equation with this interaction operator uniquely
determines the dynamics of the system. The equation of motion
(\ref{main}) with this interaction operator is well-defined and
allows one to obtain the T-matrix and the evolution operator
without resorting to regularization and renormalization. It is
easy to show that the evolution operator constructed in this way
(the evolution operator (12) with the T-matrix given by (70)) is
unitary and satisfies the composition law (\ref{law}) up to
next-to-leading order.

\section{Integral Equations}

As we have shown up to next-to-leading order of the EFT with the
Lagrangian (19), the dynamics of nucleons at low energies is
governed by the generalized dynamical equation (\ref{main}) with
the nonlocal-in-time interaction operator (\ref{gio}), provided
the parameters of the theory satisfy the condition (64). Within
the GQD this operator is well defined. Of course, the dynamical
situation in the theory with such an interaction differs from that
in the theory with ordinary potentials. The
operator$\tilde{S}(t_2,t_1)$ is only an operator-valued
generalized function on the Hilbert space. However, as we have
noted this is not at variance with the general requirements of
quantum theory, since the evolution operator given by (51) is an
operator on this space. The generalized dynamical equation
(\ref{main}) is not equivalent to the Schr{\"o}dinger equation in
this case. Nevertheless, it is a well defined equation, and allows
one to obtain the T-matrix without resorting to regularization and
renormalization procedures. This is very important for practical
calculations, since for solving realistic problems one has to deal
not only with the contact component of the NN interaction but also
with its long-range one. In this case the interaction operator is
of the form
\begin{eqnarray}
 <{\bf p}_2| {H}^{(s)}_{int}(\tau)|{\bf p}_1> =
 -2i\delta(\tau) V({\bf p}_2,{\bf p}_1)\nonumber\\+
 <{\bf p}_2| {H}_{non}(\tau)|{\bf p}_1>, \label{operator}
\end{eqnarray}
where $ <{\bf p}_2| {H}_{non}(\tau)|{\bf p}_1>$ represents the
contact nonlocal-in-time component and $V({\bf p}_2,{\bf p}_1)$ is
a  potential describing the long-range component of the NN
interaction. In general it consists of the meson-exchange
potentials and the Coulomb potential in the proton-proton channel.
In the leading order the nonlocal component is given by (53) and
for the interaction operator, we can write
\begin{eqnarray}
 <{\bf p}_2| {H}^{(s)}_{int}(\tau)|{\bf p}_1> =
 \frac{4\sqrt{\pi}\exp(i\frac{3\pi}{4})}{m^{\frac{3}{2}}\sqrt{\tau}}+\frac{16\pi^2}{m^3C_0^{(R)}}\nonumber\\
 -2i\delta(\tau)V({\bf p}_2,{\bf p}_1). \label{oper}
\end{eqnarray}
 In the case of such an interaction operator,  the solution
of the dynamical equation (16) can be represented (see Appendix A)
in the form
\begin{eqnarray}
 <{\bf p}_2| T(z)|{\bf p}_1> =t_0(z)+t_1(z;{\bf p}_1)+
 t_1(z;{\bf p}_2)\nonumber\\
 +t_2(z;{\bf p}_2,{\bf p}_1), \label{Tmatrix}
\end{eqnarray}
where $t_2(z;{\bf p}_2,{\bf p}_1)$ is a solution of the equation
\begin{eqnarray}
t_2(z;{\bf p}_2,{\bf p}_1)=V({\bf p}_2,{\bf p}_1)\nonumber\\ +\int
\frac{d^3q}{(2\pi)^3}\frac{K(z;{\bf p}_2,{\bf
q})}{z-E_q}t_2(z;{\bf q},{\bf p}_1),\label{t3}
\end{eqnarray}
with
\begin{eqnarray}
K(z;{\bf p}_2,{\bf q})=N(z)\int \frac{d^3k}{(2\pi)^3}\frac{V({\bf
p}_2,{\bf k})}{z-E_k} +V({\bf p}_2,{\bf q}),
\end{eqnarray}
and the functions $t_0(z)$ and $t_1(z;{\bf p})$ are defined as
\begin{eqnarray}
t_1(z;{\bf p})=N(z)\int \frac{d^3k}{(2\pi)^3}\frac{t_2(z;{\bf
p},{\bf q})}{z-E_k},\label{t1}
\end{eqnarray}
\begin{eqnarray}
t_0(z)=N(z)\left(1+\int \frac{d^3q}{(2\pi)^3}\frac{t_1(z;{\bf
q})}{z-E_q}\right),\label{t0}
\end{eqnarray}
where
$$N(z)=-\left(\frac{1}{C_0^{(R)}}+\frac{im^{
\frac{3}{2}}{\sqrt{z}}}{4\pi}\right)^{-1}.$$ Equations with the
next-to-leading order corrections can be derived in the same way.

 In
Weinberg's power counting the one-pion-exchange potential is of
 leading order. Hence in this order the NN interaction operator
can be expressed as
\begin{eqnarray}
 <{\bf p}_2| {H}^{(s)}_{int}(\tau)|{\bf p}_1> =
\frac{4\sqrt{\pi}\exp(i\frac{3\pi}{4})}{m^{\frac{3}{2}}\sqrt{\tau}}+
 \frac{16\pi^2}{m^{3}C_0^{(R)}}\nonumber\\
 -2i\delta(\tau)V_\pi({\bf p}_2,{\bf p}_1), \label{pioper+}
\end{eqnarray}
where $V_\pi({\bf p}_2,{\bf p}_1)$ is the conventional
one-pion-exchange potential. Substituting this potential into
Eq.(\ref{pioper+}) and solving it numerically, one can easily
obtain the T-matrix and hence the evolution operator. Note that
conventional way of solving the above problem is the formal use of
the potential
\begin{eqnarray}
V_0({\bf p}_2,{\bf p}_1)=\widetilde{C}+V_\pi({\bf p}_2,{\bf p}_1)
\end{eqnarray}
(see, for example, Refs.[7,11]). We say "formal" since the use of
such a potential leads to UV divergences, and the Schr{\"o}dinger
and LS equations require regularization and renormalization. On
the other hand, as we have shown, the contact interaction, which
in (83) is formally represented by the term $\widetilde{C}$, is
parametrized by the operator (53) (the first two terms in the
operator (82)). In this case we deal with the well defined
interaction operators and Eq.(\ref{pioper+}) which does not
require regularization and renormalization. By using Eq.(57), one
can obtain the T-matrix so easily as in the case of the pure
one-pion-exchange potential.

In order to take into account electromagnetic corrections, one has
to include the Coulomb potential into the NN interaction operator
\begin{eqnarray}
<{\bf p}_2| {H}^{(s)}_{int}(\tau)|{\bf p}_1> =
\frac{4\sqrt{\pi}\exp(i\frac{3\pi}{4})}{m^{\frac{3}{2}}\sqrt{\tau}}+
 \frac{16\pi^2}{m^{3}C_0^{(R)}}\nonumber\\
 -2i\delta(\tau)V_\pi({\bf p}_2,{\bf p}_1)
-2i\delta(\tau)\frac{e_{12}\alpha}{{\bf q}^2},\label{em}
\end{eqnarray}
where $e_{12}=1$ for the proton-proton channel and zero otherwise.
Eq.(75) with the interaction operator (\ref{em}) can be easily
solved numerically. Thus this equation allows one to investigate
electromagnetic corrections without resorting to renormalization.

\section{Summary and Discussion}

We have shown that the formalism of the GQD allows one to
formulate the effective field theory of nuclear forces  as an
internally consistent theory with equations that do not require
regularization and renormalization. It has been shown that the
effective NN interaction is nonlocal in time, and low energy
nucleon dynamics is governed by the generalized dynamical equation
(\ref{main}) with a nonlocal-in-time interaction operator. In
leading order of the EFT approach this operator is given by
(\ref{pioper}). It should be noted that the generalized
interaction operator $H_{int}(t_2,t_1)$ is not an effective
interaction Hamiltonian (only in the local case they are related
by Eq.(8)), and, in contrast with such Hamiltonians, the use of
the generalized interaction operators permits a natural
parametrization of the NN interaction: The generalized interaction
operator (\ref{pioper}) generates the unitary evolution of a
nucleon system. In Sec.III we have demonstrated the advantages of
the GQD in describing the leading order nucleon dynamics in
comparison with the ordinary methods based on the use of the
singular potential $<{\bf {p}}_2|V|{\bf {p}}_1>=C_0$. This
potential does not make sense without renormalization and hence
does not contain all the needed dynamical information. In
addition, one needs to use some empirical data as renormalization
constants.
 In contrast with this singular potential, the generalized interaction
 operator (\ref{pioper}) contains all the needed dynamical information. For example,
 it contains the observable, renormalized parameter $C_0^{(R)}$.
The dynamical equation with the interaction operator
(\ref{pioper}) is well defined and allows one to construct the
T-matrix and the evolution operator without resorting to
regularization and renormalization procedures.    The dynamical
information contains not only in the form of the operator
(\ref{pioper}) that, as has been shown, is a consequence of the
symmetries of QCD but     also in its dependence on the time
duration $\tau$ of the interaction in a system.

As has been shown, only values of $H_{int}^{(s)}(\tau)$ in the
infinitesimal neighborhood of the point $\tau=0$, i.e.  at scales
of the underlying theory are relevant. As already stated, the
operator $H_{int}^{(s)}(\tau)$ describes the contribution to the
evolution operator from the processes with infinitesimal time
duration of interaction. The above means that
\begin{eqnarray}
<\psi_2|H_{int}^{(s)}(\tau)|\psi_1>\tend\limits_{t_2\tend
t_1}<\psi_2|\tilde S_q(t_2,t_1)|\psi_1>
\end{eqnarray}
where $<\psi_2|\tilde S_q(t_2,t_1)|\psi_1>$ describe the
contributions to the evolution operator from the processes in
which the quark and gluon degrees of freedom can come to play.
Here one of the advantages of the formulation of the theory in
terms of the operators $\tilde S(t_2,t_1)$ becomes apparent: The
relevant amplitudes $<\psi_2|\tilde S(t_2,t_1)|\psi_1>$ of the
underlying theory can be directly used as the matrix elements of
the generalized interaction operator $H_{int}(t_2,t_1)$ generating
low energy dynamics. These degrees of freedom manifest themselves
through the $\tau$ dependence of the operator (\ref{pioper}). Thus
the parameter $C_0^{(R)}$ in Eq.(\ref{pioper}) parametrizes the
leading order effects of the underlying physics on low energy
dynamics, and hence can be computed in terms of parameters in QCD.
At the same time, this parameter can be determined from low energy
experiments.

The evolution operator governing the two-nucleon dynamics in
leading order is of the form (54). In Sec.IV we have shown that
the group of these operators has no infinitesimal generator and
hence the dynamics is non-Hamiltonian. This means that within the
EFT approach there are no potential satisfying the requirements of
ordinary quantum mechanics that could govern low energy nucleon
dynamics. For example, the T-matrix given by (52) does not satisfy
the LS equation and has properties that are at variance with the
Hamiltonian formalism. On the other hand, as we have seen, the
T-matrix satisfies the generalized dynamical equation (16) with
the interaction operator given by (\ref{pioper}). The essential
lesson we have learned from the above analysis is that many
problems of the EFT of nuclear forces arise because of ignoring
the fact that low energy dynamics produced by this theory is
non-Hamiltonian. For example, the above statement that there are
not any equations for renormalized amplitudes in a subtractive EFT
should mean that there are no such equations within Hamiltonian
formalism. From a more general point of view provided by the GQD
we see that such equations exist, and the theory can be formulated
in a completely consistent way. An EFT describes low energy
physics in terms of a few parameters, and these low energy
parameters can be computed in terms of a more fundamental high
energy theory. It is remarkable that the interaction operator
(\ref{pioper}) parametrizing the leading order contact component
of the NN interaction is uniquely determined by the symmetries of
QCD and contains one of these parameters $C_0^{(R)}$. Other
renormalized parameters will be contained in higher order
corrections to the generalized interaction operator, and this
corrections can be obtained order by order within the EFT
approach.     This has been demonstrated in next-to-leading order
by using the example of the EFT of short-range forces developed by
van Kolck \cite{Kolck}.

We have shown that up to next-to-leading order the dynamics of the
theory is governed by the generalized dynamical equation with the
nonlocal-in-time interaction operator (74). These results have
been obtained in the particular case where the parameters of the
theory satisfy the condition (64). This limitation is a
consequence of the fact that we used the assumption that the
two-particle T-matrix of the theory is of the form (62), while
this assumption does not fulfilled in general. The generalization
of these results to the case where the condition (64) is not
satisfied can only lead to another form of the generalized
interaction operator. The general case will be discussed in detail
in another paper. At the same time, the model we have considered
in Sec.V provide the transparent illustration of the fact that in
any order of the EFT approach low energy dynamics of nucleons is
governed by the generalized dynamical equation (\ref{main}) with a
nonlocal-in-time interaction operator. As has been proved, the
assumption that the relevant solution of Eq.(16) should be of the
form (62) uniquely determines this solution and leads to the
condition (64). In this way we arrive at exactly the same on-shell
T-matrix that has been obtained in Ref.\cite{Kolck} by summing the
bubble diagrams up to next-to-leading order.  The corresponding
generalized interaction operator is of the form (74). The
generalized dynamical equation (\ref{main}) with this interaction
operator is well defined and allows one to describe the dynamics
of the system in a consistent way.

The above method for constructing the effective interaction
operator is inapplicable in general, because the parameters of the
theory need not satisfy the condition (64). In general, in order
to construct the generalized interaction operator parametrizing
the interaction in the EFT of nuclear forces, one has to analyze
the contributions from the off-shell amplitudes to $T(z)$ in the
limit $|z|\to\infty$. In this limit the main contribution to the
T-matrix comes from the bubble diagrams including potential pions
that produce the pion-exchange potential. By summing the bubble
diagrams in the large $z$ limit, one can determine the contact
component of the interaction operator parametrizing the NN
interaction. Thus the NN interaction operator can be represented
as the sum (75) of the nonlocal contact and long-range components
that can be obtained separately. The contact component is
represented by a nonlocal-in-time interaction operator, while the
long-range one is represented by the pion-exchange potential. In
general, the long-range component should be supplemented by the
Coulomb potential. It is extremely important, that starting with
this NN interaction operator, we can construct the T-matrix and
the evolution operator without resorting to summing all relevant
diagrams and using regularization and renormalization procedures.
The generalized dynamical equation (\ref{main}) with this
interaction operator is well defined and can be reduced to
integral equations which do not require renormalization. This has
been demonstrated in leading order of the EFT approach. The
operator (76) parametrizes the interaction that includes not only
the leading order contact component but also the long-range one
being described by the potential $V({\bf p}_2,{\bf p}_1)$. The
generalized dynamical equation with the interaction operator (76)
can be reduced to the integral equation (78).  This  equation is
so convenient for numerical calculations as the LS equation which
is its particular case where the parameter $C_0^{(R)}$ tends to
zero and hence one can neglect the contact component. In this case
the theory is reduced to the ordinary theory of the NN interaction
based on the use of the pion-exchange potential.

Finally, we have shown that the use of the formalism of the GQD
allows one to formulate the effective field theory of nuclear
forces as a completely consistent theory based on the well-defined
equation of motion that does not require regularization and
renormalization. Being formulated in this way, the effective
nuclear theory permits a natural parametrization of the
interaction of nucleons by the generalized interaction operator
that be derived order by order by using the methods of subtractive
EFT's. One can hope that this operator will be able to play the
same role in nuclear physics as the Coulomb potential in quantum
mechanics of atomic phenomena. A remarkable feature of such a
formulation of the theory is that in this case all advantages of
the operator formalism of quantum mechanics can be used. The
equation of motion (\ref{main}), for example, allows one to
construct not only the S-matrix but also the evolution operator
describing the dynamics of nucleon systems. This is very
important, because the S-matrix is not everything. For example, at
finite temperature there is no S-matrix because particles cannot
get out to infinite distances from a collision without bumping
into things \cite{Weinberg}. In conclusion, it should be
emphasized that the above is not a new approach to the EFT of
nuclear forces. We have only shown that the EFT approach gives
rise to low energy nucleon dynamics which can be described, in a
natural way, only by using the generalized dynamical equation
(\ref{main}), and its use as the equation of motion describing low
energy nucleon dynamics can allow one to formulate the effective
field theory of nuclear forces as a perfectly satisfactory theory
like the quantum mechanics of atomic phenomena.
\appendix
\section{}

 Let us consider the solution of Eq.(16) in the case where the
 interaction operator is of the form (76). From (17) and (18) it
 follows that this solution can be
 represented in the form
 \begin{equation}
<{\bf {p}}_2|T(z)|{\bf {p}}_1 >= \lim \limits_{u \tend
-\infty}<{\bf {p}}_2|T_u(z)|{\bf {p}}_1 >,
\end{equation}
 where the operator $T_u(z)$ is the solution of the equation
 \begin{equation}
T_u(z)=B(u)+(u-z)B(u)G_0(u)G_0(z)T_u(z).\label{equ}
\end{equation}
Here the operator $B(z)$ is given by
\begin{eqnarray}
 <{\bf p}_2| B(z)|{\bf p}_1> =f_1(z)
 +V({\bf p}_2,{\bf p}_1),
\end{eqnarray}
with $$f_1(z)=-\frac{4\pi}{m^\frac{3}{2}\sqrt{-z}}-
 \frac{16\pi^2}{m^{3}C_0^{(R)}z}.$$
 The solution of Eq.(A2) can be represented in the form
\begin{eqnarray}
<{\bf {p}}_2|T_u(z)|{\bf {p}}_1
>=t_0^{(u)}(z)+t_1^{(u)}(z;{\bf {p}}_1)\nonumber\\+
\widetilde{t}_1^{(u)}(z;{\bf {p}}_2)+t_2^{(u)}(z;{\bf {p}}_1,{\bf
{p}}_2).
\end{eqnarray}
 Substituting this representation in Eq.(\ref{equ})
yields the following equations for $t_0^{(u)}(z)$,
$t_1^{(u)}(z;{\bf {p}})$, $\widetilde{t}_1^{(u)}(z;{\bf {p}}_2)$
and $t_2^{(u)}(z;{\bf {p}}_1,{\bf {p}}_2)$:
\begin{eqnarray}
t_0^{(u)}(z)=f_1(u)+(u-z)f_1(u)\nonumber\\
\times\int
\frac{d^3k}{(2\pi)^3}\frac{\left(t_0^{(u)}(z)+t_1^{(u)}(z;{\bf
k})\right)} {(z-E_k)(u-E_k)};\label{tu0}
\end{eqnarray}
\begin{eqnarray}
 t_1^{(u)}(z;{\bf p})=(u-z)\int
\frac{d^3k}{(2\pi)^3} \frac{\left(t_0^{(u)}(z)+t_1^{(u)}(z;{\bf
k})\right)}
{(z-E_k)(u-E_k)}\nonumber\\
\times V({\bf k},{\bf p}); \label{tu1}
\end{eqnarray}
\begin{eqnarray}
 \widetilde{t}_1^{(u)}(z;{\bf p})=(u-z)f_1(u) \nonumber\\
\times\int
\frac{d^3k}{(2\pi)^3}\frac{\left(\widetilde{t}_1^{(u)}(z;{\bf
p})+t_2^{(u)}(z;{\bf k},{\bf p})\right)} {(z-E_k)(u-E_k)};
\end{eqnarray}
\begin{eqnarray}
t_2^{(u)}(z;{\bf p}_2,{\bf p}_1)=V_\pi({\bf p}_2,{\bf
p}_1)+(u-z)\int \frac{d^3k}{(2\pi)^3}\nonumber\\
\frac{\left(\widetilde{t}_1^{(u)}(z;{\bf p}_2)+t_2^{(u)}(z;{\bf
p}_2,{\bf k})\right)} {(z-E_k)(u-E_k)}V({\bf k},{\bf p}_1).
\label{tu3}
\end{eqnarray}
 It is not difficult to verify that
 $$t_1^{(u)}(z;{\bf p})\equiv \widetilde{t}_1^{(u)}(z;{\bf p}).$$
By solving the above set of equations, one can obtain  the
functions $t_0^{(u)}(z)$, $t_1^{(u)}(z,{\bf{p}})$ and
$t_2^{(u)}(z,{\bf{p}}_2,{\bf{p}}_1)$ that in turn can be used for
constructing the T-matrix. In fact, from (A1) and (A2) it follows
that  the T-matrix can be represented in the form (\ref{Tmatrix})
where the functions $t_0(z)$, $t_1(z,{\bf{p}})$ and
$t_2(z,{\bf{p}}_2,{\bf{p}}_1)$ are given by
\begin{eqnarray}
t_0(z)=\lim_{u\to-\infty}t_0^{(u)}(z),\quad t_1(z,{\bf
{p}})=\lim_{u\to-\infty}t_1^{(u)}(z,{\bf {p}}),\nonumber
\\
t_2(z,{\bf {p}}_2,{\bf {p}}_1)=\lim_{u\to-\infty} t_2^{(u)}(z,{\bf
{p}}_2,{\bf {p}}_1).\nonumber
\end{eqnarray}
Taking into account that
$$
\int
\frac{d^3k}{(2\pi)^3}\frac{u-z}{(z-E_k)(u-E_k)}=\frac{m\sqrt{m}}{4\pi}
\left(\sqrt{-z}-\sqrt{-u}\right),
$$
Eq.(\ref{tu0}) can be rewritten in the form
\begin{eqnarray}
t_0^{(u)}(z)=-\frac{4\pi}{m^\frac{3}{2}
\sqrt{-u}}+t_0^{(u)}(z)\nonumber\\
\times\left(-\frac{4\pi}{m^\frac{3}{2}\sqrt{-u}}-\frac{16\pi^2}{m^3C_0^{(R)}u}\right)
\frac{m\sqrt{m}}{4\pi}
\left(\sqrt{-z}-\sqrt{-u}\right)\nonumber\\
-\frac{4\pi}{m^\frac{3}{2}\sqrt{-u}}\int d^3 k\frac{t_1(z,{\bf
{k}})}{z-E_k}+o(|u|^{-1/2}).\nonumber
\end{eqnarray}
Letting $u\to -\infty$ in this equation and assuming that $V({\bf
p}_2,{\bf p}_1)$ satisfies the ordinary requirements of quantum
mechanics,  one can easily get Eq.(\ref{t0}). In the same way,
from (\ref{tu1}) and (\ref{tu3}) one can derive Eqs.(\ref{t3}) and
(\ref{t1}).

\acknowledgments

The work was supported by Fund NIOKR of RT and Academy of Sciences
of Republic of Tatarstan N 14-98.


\begin{references}
\bibitem{EFT1}
S. Weinberg, Physica A\ {\bf 96} 327 (1979).
\bibitem{EFT2}
E. Witten, Nucl.\ Phys. B\ {\bf 122} 109 (1977).
\bibitem{EFT3}
S. Weinberg, Phys.\ Lett. B\ {\bf 251} 288 (1990); Nucl.\ Phys.\
B\ {\bf 363} 3 (1991).
\bibitem{Ordonez}
C. Ord{\`o}{\~n}ez, and U. van Kolck, Phys.\ Lett. B\ {\bf 291}
459 (1992).
\bibitem{vKolck}
U. van Kolck, Phys.\ Rev. C\ {\bf 49} 2932 (1994).
\bibitem{Ray}
C. Ord{\`o}{\~n}ez, L. Ray, and U. van Kolck, Phys.\ Rev.\ Lett.\
{\bf 72} 1982 (1994).
\bibitem{Kaplan}
D.B. Kaplan, M.J. Savage, and M.B. Wise, Phys.\ Lett. B\ {\bf 424}
390 (1998); Nucl.\ Phys. B\ {\bf 534} 329 (1998).
\bibitem{Bedaque}
P.F. Bedaque, M.J. Savage, R. Seki, U. van Kolck (Eds.), Nuclear
Physics with Effective Field Theory II, (World Scientific,
Singapore, 1999); R. Seki, U. van Kolck, M.J. Savage (Eds.),
Nuclear Physics with Effective Field Theory, (Word Scientific,
Singapore, 1998).
\bibitem{9}
U. van Kolck, Prog.\ Part.\ Nucl.\ Phys.\ {\bf 43}, 409 (1999).
\bibitem{Lepage}
G.P. Lepage, What is Renormalization?, (Word Scientific,
Singapore, 1990); nucl-th/9706029.
\bibitem{Gegelia}
J. Gegelia, Phys.\ Lett. B\ {\bf 463} 133 (1999).
\bibitem{R.Kh.:1999} R.Kh. Gainutdinov,
J. Phys. A\ {\bf 32}, 5657 (1999).
\bibitem{Feynman:1948}
R.P. Feynman, Rev.\ Mod.\ Phys.\ {\bf 20}, 367 (1948); R.P.
Feynman  and A.R. Hibbs, Quantum Mechanics and Path Integrals,
(McGraw-Hill, New York, 1965).
\bibitem{Kolck}
U. van Kolck, Nucl.\ Phys. A\ {\bf 645}, 273 (1999)
\bibitem{C}
C. Ord{\`o}{\~n}ez, L. Ray, and U. van Kolck, Phys.\ Rev. C {\bf
53} 2086 (1996).
\bibitem{R.Kh./A.A.:1999} R.Kh. Gainutdinov and  A.A. Mutygullina,
Yad.\ Fiz.\ {\bf 62}, 2061 (1999).
 \bibitem{R.Kh.:2001} R. Kh.
Gainutdinov, hep-th/0107139.
 \bibitem{asym}
 F. W. J. Olver,
Introduction to Asymptotics and Special Functions, (Academic
Press, New York and London, 1974).
 \bibitem{Weinberg}
S. Weinberg, hep-th/9702027.
\end{references}
\end{document}